\documentclass{naturesupplemental}
\usepackage{graphicx}
\usepackage{graphics}
\usepackage{amsfonts}
\usepackage{amsmath}
\usepackage{epsfig}

\title{Direct Observation of the Superfluid Phase Transition in Ultracold Fermi Gases}

\author{Martin W. Zwierlein, Christian H. Schunck, Andr\'e Schirotzek, and Wolfgang Ketterle}

\begin{document}
\catcode`\ä = \active \catcode`\ö = \active \catcode`\ü = \active
\catcode`\Ä = \active \catcode`\Ö = \active \catcode`\Ü = \active
\catcode`\ß = \active \catcode`\é = \active \catcode`\è = \active
\catcode`\ë = \active \catcode`\ô = \active \catcode`\ê = \active
\catcode`\ø = \active \catcode`\ò = \active \catcode`\í = \active
\catcode`\Ó = \active \catcode`\ú = \active \catcode`\á = \active
\catcode`\ã = \active
\defä{\"a} \defö{\"o} \defü{\"u} \defÄ{\"A} \defÖ{\"O} \defÜ{\"U} \defß{\ss} \defé{\'{e}}
\defè{\`{e}} \defë{\"{e}} \defô{\^{o}} \defê{\^{e}} \defø{\o} \defò{\`{o}} \defí{\'{i}}
\defÓ{\'{O}} \defú{\'{u}} \defá{\'{a}} \defã{\~{a}}
\newcommand{\li}{$^6$Li}
\newcommand{\na}{$^{23}$Na}
\newcommand{\cs}{$^{133}$Cs}
\newcommand{\kk}{$^{40}$K}
\newcommand{\rb}{$^{87}$Rb}
\newcommand{\vect}[1]{\mathbf #1}
\newcommand{\g}{g^{(2)}}
\newcommand{\one}{$\left|\uparrow\right>$}
\newcommand{\two}{$\left|\downarrow\right>$}
\newcommand{\V}{V_{12}}
\newcommand{\kfa}{\frac{1}{k_F a}}
\maketitle

\begin{affiliations}
 \spacing{1}
 \item Department of Physics, MIT-Harvard Center for Ultracold Atoms, and Research Laboratory of Electronics,
 MIT, Cambridge, MA 02139
\end{affiliations}

\begin{abstract}
Water freezes into ice, atomic spins spontaneously
align in a magnet, liquid helium becomes superfluid: Phase
transitions are dramatic phenomena. However, despite the drastic
change in the system's behaviour, observing the transition can
sometimes be subtle. The hallmark of Bose-Einstein condensation
(BEC) and superfluidity in trapped, weakly interacting Bose gases is the
sudden appearance of a dense central core inside a thermal
cloud~\cite{ande95,davi95bec,grei03mol_bec,zwie03molBEC,bart04,bour04coll,Part05}.
In strongly interacting gases, such as the recently observed
fermionic superfluids~\cite{zwie05Vort}, this clear separation
between the superfluid and the normal parts of the cloud is no
longer given. Condensates of fermion pairs could be detected only
using magnetic field sweeps~\cite{rega04,zwie04rescond,zwie04form}
into the weakly interacting regime. The quantitative description of these sweeps presents a major theoretical challenge.
Here we demonstrate that the superfluid phase transition can be
directly observed by sudden changes in the shape of the clouds, in
complete analogy to the case of weakly interacting Bose gases. By
preparing unequal mixtures of the two spin components involved in
the pairing~\cite{zwie05imbalance,part06phase}, we greatly enhance the contrast between the
superfluid core and the normal component. Furthermore, the
non-interacting wings of excess atoms serve as a direct and
reliable thermometer. Even in the normal state, strong
interactions significantly deform the density profile of the
majority spin component. We show that it is these interactions
which drive the normal-to-superfluid transition at the critical
population imbalance of 70(5)\%~\cite{zwie05imbalance}.
\end{abstract}

The dramatic signature of BEC in weakly interacting gases in atom
traps derives from a natural hierarchy of energy scales: The
critical temperature for condensation $T_C \propto n^{2/3}$ at
particle density $n$ is much larger than the chemical potential (divided by the Boltzmann constant $k_B$) of
a pure condensate, $\mu \propto n a$, which measures the interaction strength
between atoms ($a$ is the scattering length). Hence, for weak (repulsive) interactions ($n a^3 \ll 1$),
the condensate is clearly distinguished from the cloud of
uncondensed atoms through its smaller size and higher density.
However, as the interactions are increased, for example by tuning
$a$ using a Feshbach
resonance, this hierarchy
of energy scales breaks down, as $\mu$ can now become comparable
to $k_B T_C$. In Fermi gases with weak attractive interaction ($a < 0$),
$\mu \propto E_F$ will even exceed $k_B T_C \propto E_F \exp^{-\pi / 2
k_F \left|a\right|}$ by far ($E_F = \hbar^2 k_F^2 / 2 m$ is the
Fermi energy). Both the normal and the condensed cloud will here
be of the same size given by the Fermi Radius $R_F \propto
\sqrt{E_F}$.

The phase transition from the normal to the superfluid state,
although dramatic in its consequences, is thus not revealed by a
major change in the appearance of the gas. Indeed, in strongly interacting Fermi gases no deviation from a normal cloud's shape could so far be detected; neither in the unitary regime, where $a$ diverges, nor on the attractive (BCS-)side of a Feshbach resonance. Theoretical works predicted small
"kinks"~\cite{chio02,ho04uni,pera04temp} or other slight deviations~\cite{staj05dens} in the
density profiles of the gas in the superfluid regime, but after
line-of-sight integration these effects have so far been too small
to be observable. Condensates could only be observed via rapid
magnetic field ramps to the BEC-side ($a > 0$) of the Feshbach
resonance, performed during expansion~\cite{rega04,zwie04rescond}.
This suddenly reduced the condensate's chemical potential and let
the thermal fraction grow beyond the condensate size. A similar
ramp was used to detect vortices on resonance
and on the BCS-side in the demonstration of fermionic
superfluidity~\cite{zwie05Vort}. However, these magnetic field
ramps are difficult to model theoretically, and a satisfactory
quantitative comparison of e.g. the condensate fraction with
experiments has not been accomplished~\cite{ho04proj,pera05,altm05proj,chen05phase}.

In this work we demonstrate that the normal-to-superfluid phase
transition in a strongly interacting Fermi gas can be directly
observed from  absorption profiles, without the need of any
magnetic field ramps. As in the case of weakly interacting BECs,
preparation, expansion and detection of the sample all take place
at the same, fixed magnetic field and scattering length. As
for BECs, the phase transition is observed by a sudden change in
the shape of the cloud during time-of-flight expansion, when the
trap depth is decreased below a critical value. To clearly
distinguish the superfluid from the normal component we break the number symmetry
between spin up (majority atom number $N_\uparrow$) and spin down (minority atom number $N_\downarrow$) and
produce an unequal mixture of fermions (imbalance parameter $\delta = (N_\uparrow - N_\downarrow)/(N_\uparrow + N_\downarrow)$). Standard BCS
superfluidity requires equal densities of the two spin components.
Hence, when cooled below the phase transition the cloud should
show a sudden onset of a superfluid region of equal densities.
Indeed, below a critical temperature, we observe a bimodal density
distribution of the smaller cloud.

Breaking the symmetry in atom numbers thus produces a direct and
striking signature of the superfluid phase transition.  A similar
situation has been encountered in Bose-Einstein condensation,
where breaking the symmetry of a spherical trap resulted in
dramatic anisotropic expansion of the condensate, now a hallmark
of the BEC phase transition.

\begin{figure}
    \begin{center}
    \includegraphics[width=6in]{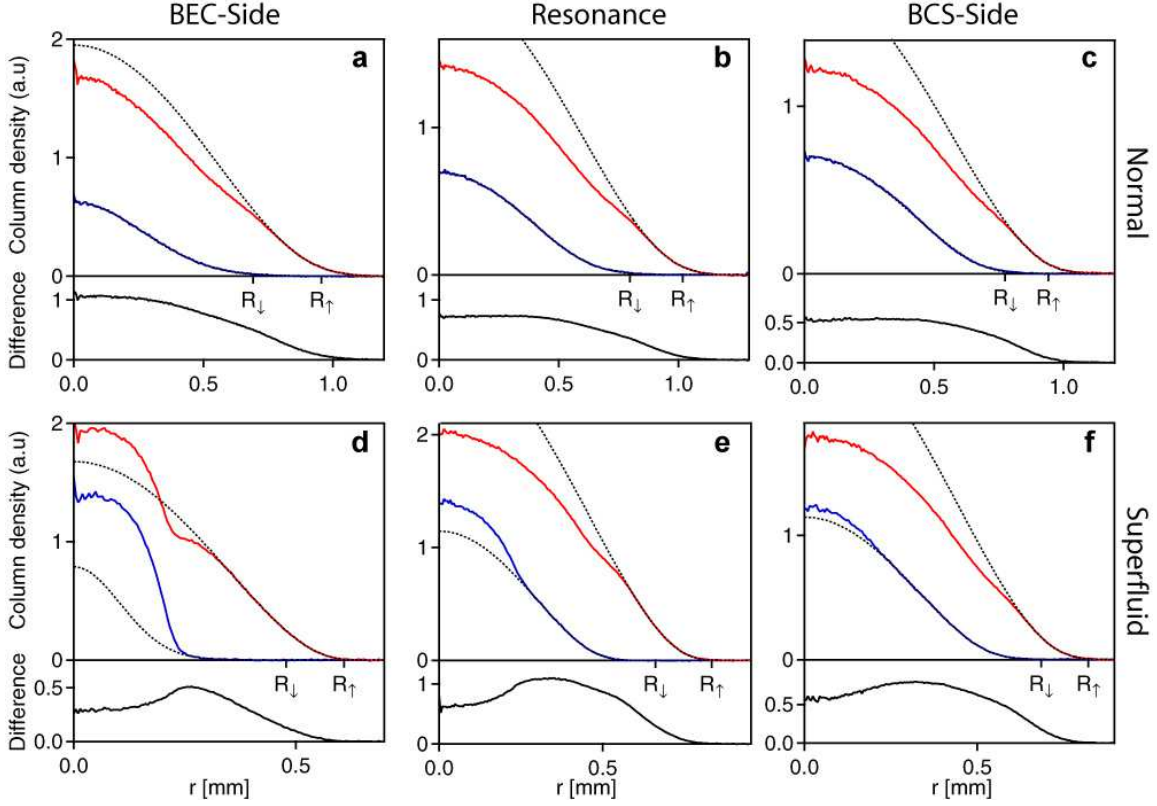}
    \caption[Title]{(Color online) Direct observation of the phase transition in a strongly interacting two-state mixture of fermions with imbalanced spin populations. Top {\bf a}-{\bf c} and bottom {\bf d}-{\bf f} rows show the normal and the superfluid state, respectively. Panels {\bf a} and {\bf d} were obtained in the BEC-regime (at 781 G), {\bf b},{\bf e} on resonance (B = 834 G) and {\bf c},{\bf f} on the BCS-side of the Feshbach resonance (at 853 G). The profiles represent the azimuthal average of the column density after 10 ms (BEC-side) or 11 ms (resonance and BCS-side) of expansion. The appearance of a dense central feature in the smaller component marks the onset of condensation. The condensate causes a clear depletion in the difference profiles (bottom of each panel). Both in the normal and in the superfluid state, interactions between the two spin states are manifest in the strong deformation of the larger component. The dashed lines show Thomas-Fermi fits to the wings of the column density. The radii $R_\uparrow$ and $R_\downarrow$ mark the Fermi radius of a ballistically expanding, non-interacting cloud with atom number $N_\uparrow$, $N_\downarrow$. The trap depth $U$, the atom numbers, the population imbalance $\delta = (N_\uparrow - N_\downarrow)/(N_\uparrow + N_\downarrow)$, the interaction parameter $1/k_F a$, the temperature $T$ and the reduced temperature $T/T_F$ were: {\bf a}, $U = 4.8 \,\mu\rm K$, $N_\uparrow = 1.8 \times 10^7$, $N_\downarrow = 2.6 \times 10^6$, $\delta = 75\%$, $1/k_F a = 0.42$, $T = 350 \,\rm nK$, $T/T_F = 0.20$. {\bf b}, $U = 3.2 \,\mu\rm K$, $N_\uparrow = 1.8 \times 10^7$, $N_\downarrow = 4.2 \times 10^6$, $\delta = 63\%$, $1/k_F a = 0$ (resonance), $T = 260 \,\rm nK$, $T/T_F = 0.15$. {\bf c}, $U = 2.5 \,\mu\rm K$, $N_\uparrow = 1.5 \times 10^7$, $N_\downarrow = 4.5 \times 10^6$, $\delta = 52\%$, $1/k_F a = -0.13$, $T = 190 \,\rm nK$, $T/T_F = 0.12$. {\bf d}, $U = 0.8 \,\mu\rm K$, $N_\uparrow = 6.5 \times 10^6$, $N_\downarrow = 1.5 \times 10^6$, $\delta = 62\%$, $1/k_F a = 0.67$, $T = 50 \,\rm nK$, $T/T_F \le 0.05$.  
{\bf e}, $U = 1.1 \,\mu\rm K$, $N_\uparrow = 1.5 \times 10^7$, $N_\downarrow = 3.8 \times 10^6$, $\delta = 60\%$, $1/k_F a = 0$ (resonance), $T = 70 \,\rm nK$, $T/T_F = 0.06$. {\bf f}, $U = 1.2 \,\mu\rm K$, $N_\uparrow = 1.3 \times 10^7$, $N_\downarrow = 4.4 \times 10^6$, $\delta = 50\%$, $1/k_F a = -0.15$, $T = 100 \,\rm nK$, $T/T_F = 0.08$.} \label{fig:densityprofs}
    \end{center}
\end{figure}

Fig.~1 shows column density profiles of the
two imbalanced spin states for different points along the evaporation path
corresponding to different temperatures, and for three
magnetic fields corresponding to the BEC-side, exact resonance and
the BCS-side of the resonance.
For high final trap depths (upper panels in Fig.~1), the smaller cloud has the expected shape of a normal, non-superfluid gas: It is very well fit using a single, finite temperature Thomas-Fermi-profile (with central optical density, radius and the fugacity as independent fit-parameters). However, below a critical trap depth, a second, denser feature appears in the center of the minority component (lower panels in Fig.~1). This onset of bimodality occurs very suddenly as the trap depth is lowered, as can be seen from Fig.~2: Around the critical point, the atom number (Fig.~2a) and population imbalance (Fig.~2b) are practically constant, and the temperature (Fig.~2c) varies in a smooth linear way with the trap depth. In contrast, below the critical trap depth, the shape of the smaller cloud starts to deviate drastically from the Thomas-Fermi distribution of a normal gas, as quantified in Fig.~2d. This
sudden increase in the standard deviation of a fit to a
single-component fitting function is a standard way of identifying
the BEC phase transition in a model-independent way~\cite{davi95bec}.

\begin{figure}
    \begin{center}
    \includegraphics[width=4.5in]{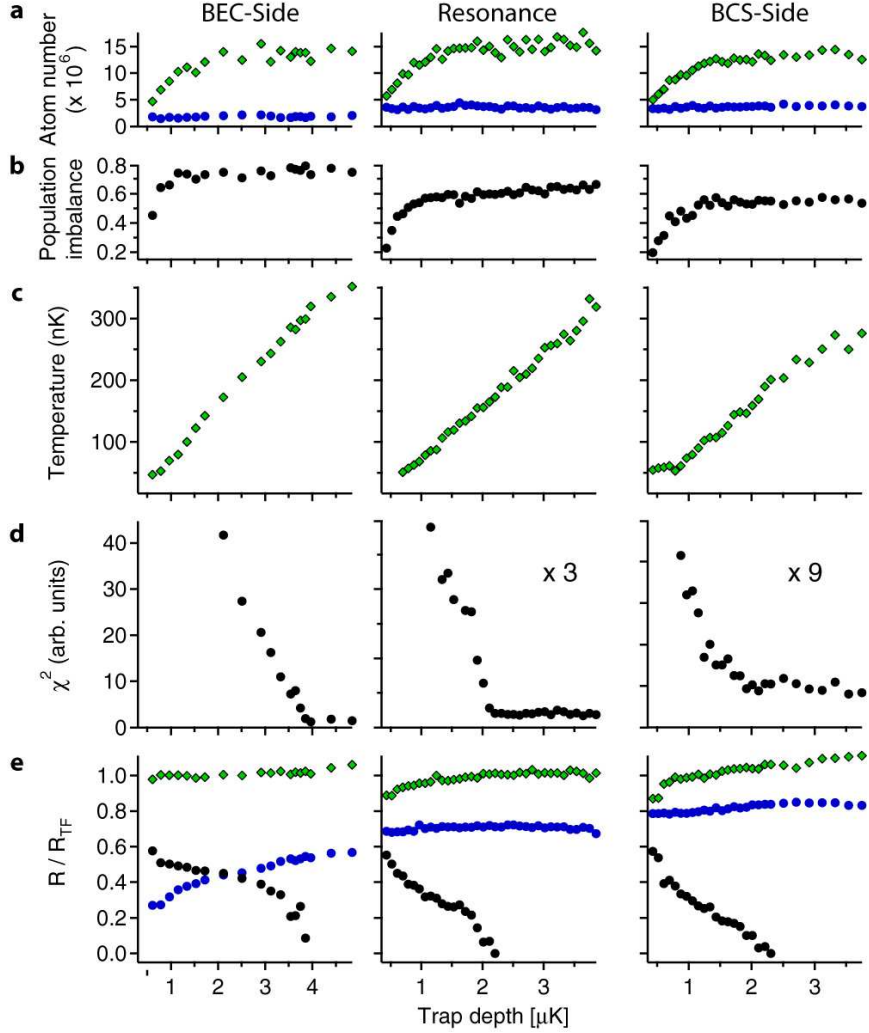}
    \caption[Title]{(Color online) Characterization of the phase
transition. The data characterize the evolution of the fermion
mixture as the cloud is evaporatively cooled by lowering the trap depth. The chosen magnetic fields are identical to those in Fig.~1. Data obtained from the
majority (minority) cloud are shown as diamonds (circles). Shown
are a) the atom number, b) population imbalance between the two
spin states and c) the temperature of the spin mixture as
determined from the non-interacting wings of the larger cloud's
profile. d) A finite temperature Fermi-Dirac (for resonance and
the BCS-side) or gaussian (for the BEC-side) distribution is fit
to the minority cloud. The phase transition is marked by a sudden
increase in $\chi^2$ as the condensate starts to appear. e) Outer
radii of the majority and minority cloud (for the minority cloud
on the BEC-side: thermal cloud radius, all other cases: Thomas-Fermi radius) as well as the condensate
radius, defined as the position of the "kink" in the minority profile (see
Fig.~1). The radii of the majority and minority
clouds are normalized to the Fermi-Radii $R_{\uparrow,\downarrow}$ of
non-interacting atoms with atom number $N_{\uparrow,\downarrow}$, and adjusted for
ballistic (hydrodynamic) expansion. Note that the imbalance decreases during evaporation because the larger majority cloud incurs stronger evaporative losses. For the data, three (BEC\&Resonance) to five (BCS) independent measurements were averaged.} \label{fig:phasetransition}
    \end{center}
\end{figure}

Fig.~2e displays the fact that below the
critical trap depth a new, third radius is required to describe
the two clouds. As we will see below, the appearance of this
central feature coincides with the appearance of the fermion pair
condensate in experiments involving the magnetic field ramp
technique~\cite{zwie04rescond,zwie04form,zwie05imbalance}. It is this condensate which
contains the superfluid vortices in~\cite{zwie05Vort,zwie05imbalance}. We are thus naturally led to
interpret the central core as the condensate of fermion pairs, and
the outer wings as the normal, uncondensed part of the cloud. This
constitutes the first direct observation of the
normal-to-superfluid phase transition in resonantly interacting
Fermi gases on resonance and on the BCS-side (i.e. without a
magnetic field sweep that so far cannot be quantitatively accounted for). 

Already at high temperatures, above the phase transition, the
larger cloud's profile is strongly deformed in the presence of the
smaller cloud, a direct signature of interaction. Indeed, on
resonance the cloud size of the minority component is significantly
smaller than that of a non-interacting sample with the same
number of atoms (see Fig.~2e). At the
phase transition, the outer radii of the clouds do not change
abruptly. This demonstrates that interactions, not superfluidity,
are the main mechanism behind the reduced cloud size of an
interacting Fermi gas.

On the BEC-side, the condensate is clearly visible in the larger
cloud. On resonance, however, the condensate is not easily
discernible in the larger component's profiles at the scale of
Fig.~1. Nevertheless, we have found a very
faint but reproducible trace of the condensate when analyzing the
curvature of these column density profiles (see Fig.~S1 in the Supplementary Information).
On resonance and on the BCS-side, the onset of bimodality in the smaller cloud can be clearly observed for imbalances larger than $\sim 20\%$ (but below a certain critical imbalance, see below), for which the condensate is small compared to the minority cloud size. With increasing magnetic field on the BCS-side (i.e. with decreasing interaction strength), the bimodality becomes less pronounced and is not clearly discerned beyond 853 G ($1/k_F a < -0.15$).

Thermometry of strongly interacting Fermi gases has always been a
major difficulty in experiments on strongly interacting
fermions~\cite{kina05heat}. A thermometer can only be reliable if
the working substance is not affected by the sample to be
measured. In equal mixtures of fermions, the two overlapping
atomic clouds are strongly interacting throughout. Temperatures
determined from a non-interacting Thomas-Fermi fit to these clouds
need calibration based on approximate theoretical calculations~\cite{kina05heat}. In addition, as will be reported elsewhere, we find that those fits do not describe the profiles of a partially superfluid Fermi gas as well as they do in the normal state, in agreement with theory~\cite{chio02,ho04uni,pera04temp,staj05dens}. 
In the case of imbalanced mixtures, the wings of the larger
component, where the spin down species are absent, are non-interacting and thus serve as a direct thermometer (see
Fig.~2c). For an imbalance of $\delta =
75(3)\%$, we determine the critical temperature for the phase
transition on the BEC-side at $1/k_F a = 0.46$ to be $T/T_F =
0.18(3)$ ($k_B T_F = \hbar \omega (3 (N_\uparrow + N_\downarrow))^{1/3}$ - Fermi energy of a non-interacting, equal mixture with the same total number of fermions $N_\uparrow + N_\downarrow$, $\omega$ - geometric mean of the trapping frequencies).
This corresponds to $T/T_{C,\downarrow} = 0.55(9)$ when comparing
the temperature to the critical temperature $T_{C,\downarrow}$ for
Bose-condensation in a non-interacting gas with $N_\downarrow$ bosons. The
reduction in the critical temperature is a direct consequence of
strong repulsive interactions between the molecules. On resonance, at $\delta = 59(3)\%$, we find
$T/T_F = 0.12(2)$, and on the BCS-side $1/k_F a = -0.14$ for
$\delta = 53(3)\%$ we obtain $T/T_F = 0.11(2)$, where we have
normalized the temperature by the Fermi temperature of an equal
mixture with the same number of atoms, $k_B T_F = \hbar \omega (3 (N_\uparrow + N_\downarrow))^{1/3}$ ($\omega/2\pi$ - geometric mean of the trapping frequencies). These are the first
directly measured and reliable temperatures for the superfluid
transition in strongly interacting Fermi gases.  They may serve as
a checkpoint for theoretical models.

We note that the critical temperature will in general depend on the population imbalance. For example, for large enough imbalance on resonance or on the BCS-side, no condensate will form even at zero temperature~\cite{zwie05imbalance}, as we discuss below. Here, the critical temperature for superfluidity will be zero.

An important qualitative difference distinguishes the BEC-side
from resonance: At the lowest temperatures on the BEC-side, the
gas consists of only two parts: The superfluid core surrounded by
a fully polarized degenerate Fermi gas of the excess species. On
resonance and on the BCS-side, however, there exists a third
region, a normal state in which both species are mixed. 
Several recent theories describe density profiles of imbalanced Fermi mixtures~\cite{pier05dens,kinn05imbal,desi06dens,yi06dens,chev06dens,haqu06dens}. Mean-field theories that neglect interactions in the normal cloud and between the normal and condensed cloud, are only in qualitative agreement with our results. Descriptions which exclude the mixed region or find superfluidity on resonance at all population imbalances are ruled out by our observations.

\begin{figure}
    \begin{center}
    \includegraphics[width=4in]{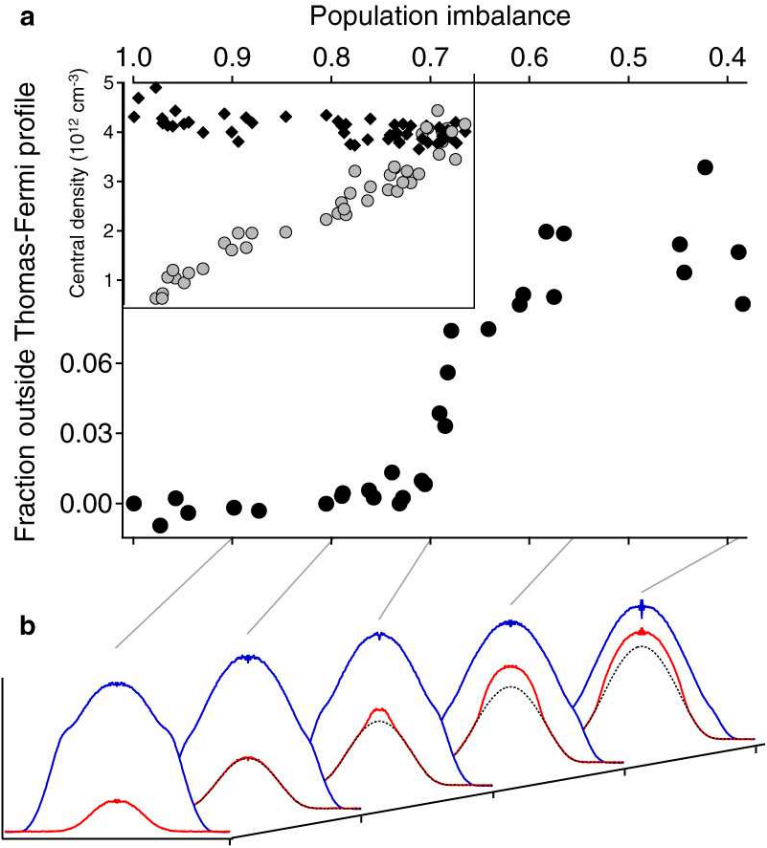}
    \caption[Title]{(Color online) Quantum phase transition to
superfluidity for decreasing population imbalance. {\bf a}, The "condensate fraction" of excess minority atoms, not contained in the Thomas-Fermi-fit, versus population imbalance on resonance. {\bf b}, Column density profiles, azimuthally averaged, for varying population imbalance. The condensate is clearly visible in the minority component as the dense central feature on top of the normal background (finite-temperature Thomas-Fermi fit, dotted lines). Below the critical imbalance $\delta_c = 70\%$, the condensate
starts to form. The inset in {\bf a} shows the central densities of the
larger (black diamonds) and smaller (grey circles) cloud in the
normal state above $\delta_c$. This demonstrates that here the
central densities are unequal, suppressing superfluidity.
The densities were calculated from the central optical density and
the fitted size of the clouds, assuming local density
approximation and adjusting for ballistic (hydrodynamic) expansion of the outer
radii of majority (minority) clouds. The data-points for the
condensate fraction show the average of several independent
measurements.} \label{fig:condfrac}
    \end{center}
\end{figure}

To elucidate the origin of the clear separation between condensate
and normal components, we varied the population imbalance at our
coldest temperatures and on resonance. Fig.~3b
shows several resulting profiles after 11 ms expansion from the
trap. For large imbalances, $\delta > 70\%$, the minority cloud is
not bimodal and well fit by a (unconstrained) Thomas-Fermi
profile. At a critical imbalance of $\delta \approx 70\%$, the
condensate appears and  then grows further as the imbalance is
reduced (for the cloud radii see Fig.~S2 in the Supplementary Information).

To characterize the appearance of the condensate for imbalances
around $\delta = 70\%$, a Thomas-Fermi profile is fit to the wings
of the minority cloud. The fraction of atoms not contained in this
fit is a measure of the condensate fraction (see
Fig.~3). We find a critical imbalance of
$\delta_c = 70(5)\%$ above which the condensate disappears. This
agrees with our previous work~\cite{zwie05imbalance}, where we
employed a rapid ramp method to the BEC-side to extract the
condensate fraction. We observed the quantum phase transition from
the superfluid to the normal state as a critical population
imbalance of $\delta_c = 70\%$ was exceeded. This strongly suggests that
the bimodality observed here directly in the minority component
and the bimodality observed in molecular clouds after a magnetic
field sweep are signatures of the same phase transition.

The transition at $\delta_c$ is known as the Clogston
limit of superfluidity~\cite{clog62,zwie05imbalance} and occurs when the chemical
potential difference $\delta\mu$ becomes larger than the (local)
superfluid gap $\Delta(\vect{r})$ (see Supplementary Information). Here we present a simple picture for the character of
this phase transition in a harmonic trap. Thomas-Fermi-fits for the normal clouds beyond $\delta_c$ allow a
simple estimate of the central 3D-density of the gas (with an estimated accuracy of 20\% for the relative density difference), shown in the inset of Fig.~3. For large imbalances, we find that
the 3D-densities differ significantly, as is expected for two
weakly interacting Fermi clouds. As the imbalance is reduced
towards the critical $\delta_c$, the central densities approach
each other and become approximately equal around $\delta_c$. This
is a direct consequence of strong interactions in the normal
state. In a non-interacting Fermi mixture with an imbalance of
$\delta_c$, the central densities would differ by a factor of 2.4.

This observation now offers an intriguing insight into the nature
of a fermionic superfluid on resonance or on the BCS-side. Already
in the normal state above $T_C$ or beyond $\delta = \delta_c$,
interactions between the two spin states are strong. Indeed, this
is directly seen in the deformation of the majority cloud due to
the presence of the minority species (see
Figs.~1,~3). However, here
these interactions are not strong enough to let the central densities of
the two clouds become comparable. At the critical imbalance the Clogston
criterion $\delta\mu = \Delta(\vect{0})$ is fulfilled in the center of the
trap. For smaller imbalance, a central superfluid region can form, the condensate. Its borders are defined
by $\delta\mu < \Delta(\vect{r})$. The simple density estimate in Fig.~\ref{fig:condfrac} suggests that this region will be of equal densities, although more refined techniques to measure small density differences have to be developed to finally conclude on this question. Outside the superfluid region there is
still a normal state with unequal densities of minority and
majority components. The discontinuity in the clouds' densities at
the normal-to-superfluid phase boundary gives rise to the visible
kink in the column density profiles. Such a density discontinuity is characteristic for a first-order phase transition.

Interestingly, most of the ``work" to build the superfluid state
is already done in the normal component by decreasing the
density difference. Consequently, the critical population
difference to form the {\it superfluid} is largely determined by the
interactions in the {\it normal} gas.

In conclusion, we have observed the normal-to-superfluid phase
transition through the direct observation of condensation in an
imbalanced Fermi mixture, on the BEC-side, on the BCS-side and
right on the Feshbach resonance. Unequal mixtures offer a direct
method of thermometry by analyzing the non-interacting wings of
the majority species. Strong interactions are already visible in
the normal cloud as marked deformations of the majority profile.
It is these interactions in the normal gas which squeeze the two
components and eventually, at the critical imbalance, let them reach almost equal densities in the center, aiding the formation of the superfluid. Our method of direct detection
of the condensate is a powerful new tool to characterize the
superfluid phase transition. At the current level of precision, the appearance of a condensate after magnetic field sweeps and the direct observation of the central dense core occur together and indicate the normal-to-superfluid phase transition. An intriguing question is whether further phases are possible, including a more exotic superfluid state with unequal densities. Several theories predict that the FFLO-state, a superfluid state
with oscillating order parameter, should be present for imbalanced
spin populations~\cite{mizu05fflo,kinn05imbal}.

\begin{methods}
Our experimental setup is described in previous
publications~\cite{zwie05Vort,zwie05imbalance}. A spin-polarized
cloud of \li\ fermions is cooled to degeneracy using a combination
of laser cooling and sympathetic cooling with sodium atoms in a
magnetic trap. After transfer into an optical trap, a variable
spin mixture of the lowest two hyperfine states, labelled \one\
and \two\, is prepared at a magnetic bias field of 875 G.
Interactions between the two spin states can be freely tuned via a
300 G wide Feshbach resonance located at $B_0 = 834\,\rm G$. At fields
below $B_0$, two-body physics supports a stable molecular bound
state (BEC-side), while at higher fields (BCS-side), no such bound
state exists for two isolated atoms. Our trap combines a magnetic
saddle potential with a weakly focused (waist $w \approx 120\, \mu
\rm m$) infrared laser beam (wavelength $\lambda = 1064\, \rm nm$),
leading to a harmonic axial confinement with oscillation frequency
of $\nu_z = 22.8(0.2) \,\rm Hz$ and a gaussian radial potential with
variable trapping frequency $\nu_r$ in the central harmonic
region. The trap depth $U$ is related to $\nu_r$ and $\nu_z$ via
$$U = \frac{1}{4}\,m (2 \pi \nu_r)^2 w^2 \left(1 - \frac{\nu_z^2}{2\nu_r^2}
\ln\left(\frac{2(\nu_r^2+\nu_z^2/2)}{\nu_z^2}\right)\right).$$

The initial degeneracy of the spin-mixture is about $T/T_F \approx
0.3$. The
strongly interacting gas is further cooled by decreasing the laser
power of the optical trap in several seconds and evaporating the
most energetic particles. During the first few seconds, the
magnetic field is adiabatically ramped to a chosen final field in
the resonance region where the last stage of the evaporation
(shown in Fig.~2) takes place. For detection, the optical trap is
switched off and the gas expands in the remaining magnetic saddle
point potential. After a variable time-of-flight an absorption
image of atoms either in state \one\ or \two\ is taken along the axial direction of the trap (the direction of the optical trapping beam). The cloud's radial symmetry allows for azimuthal averaging of the resulting column densities, leading to low-noise profiles~\cite{zwie05imbalance}.

For preparing clouds at the coldest temperatures (as shown in Fig.~3) with varying population imbalance, the spin mixture is
evaporated down to a trap depth of 1 $\mu$K over several seconds
on resonance, after which the trap depth is increased again to 1.4
$\mu$K for more harmonic confinement (trap frequencies: $\nu_r =
115(10)$ Hz and $\nu_z = 22.8(0.2)$ Hz). The temperature of the gas is
determined to be $T/T_F \le 0.06$ for all $\delta > 15\%$, and
appeared to smoothly raise to $T/T_F = 0.11$ for an equal mixture,
although thermometry in the interacting wings is problematic. The
total atom number was $1.5 \times 10^7$ and constant to within
$15\%$ for all values of $\delta$.

The error in the critical temperature $T_C/T_F$ for the phase transition is dominated by the uncertainty in the atom number entering the determination of $T_F$, which we estimate to be $30\%$~\cite{zwie05imbalance}. For $T_F$ we use the harmonic approximation for the radially gaussian trapping potential, with the measured trapping frequencies reflecting the average curvature of the gaussian potential. The phase transition is observed above $U = 2\,\mu\rm K$, where anharmonicities contribute only $3\%$ to the error in $T_F$.
Note that anharmonicities do not affect the temperature measurement performed on the majority wings: Ballistic expansion of non-interacting atoms reveals their momentum distribution, regardless of the shape of the trap.
\end{methods}

\begin{addendum}
\item We would like to thank the participants of the Aspen winter
conference on strongly interacting fermions for stimulating
discussions. This work was supported by the NSF, ONR, and
NASA.
\item[Author Information] Correspondence and requests for materials
should be addressed to M.~W.~Z.~(email: zwierlei@mit.edu).
\end{addendum}

\clearpage

\def\figurename{{\bf Supplementary Figure}}

\section*{{\large Supplementary Information}}

\section*{Supplementary Figures}

\begin{figure*}[h]
    \begin{center}
    \includegraphics[width=4.5in]{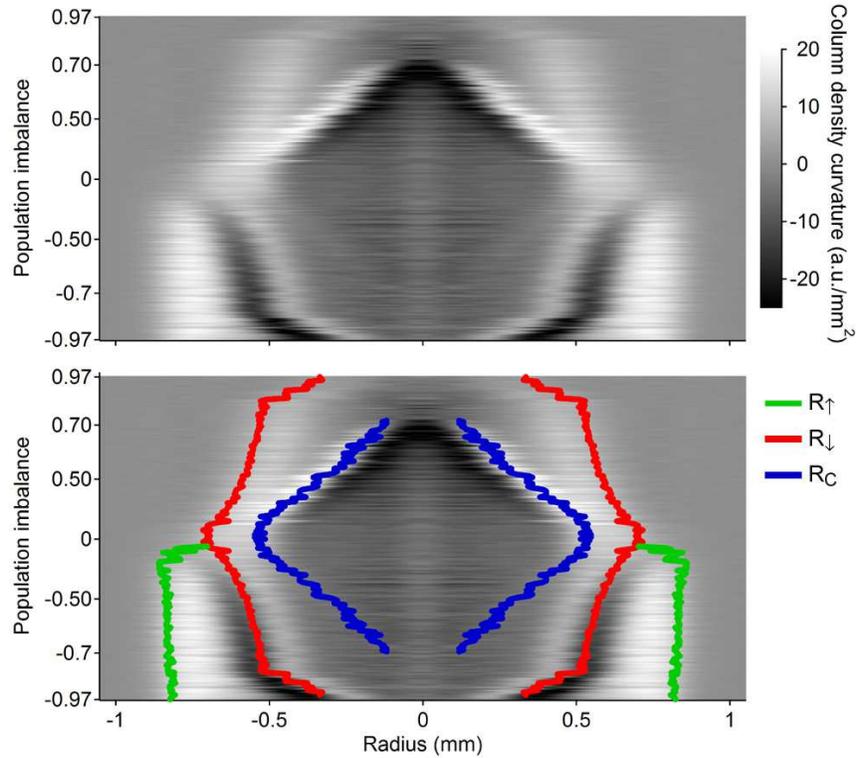}
    \caption[Title]{(Color online) Signatures of the condensate on resonance in the spatial profiles. The curvature of the observed column density is encoded in shades of gray with white (black) corresponding to positive (negative) curvature. The outer radii of the two components and the condensate radius are shown as an overlay in the lower panel. As a direct consequence of strong interactions, the minority component causes a pronounced bulge in the majority density that is reflected in the rapid variation of the profile's curvature. The condensate is clearly visible in the minority component ($\delta > 0$), but also leaves a faint trace in the majority component ($\delta < 0$). The image was composed out of 216 individual azimuthally averaged column density profiles, smoothed to reduce technical noise. Data close to the cloud's center suffer from larger noise due to the lower number of averaged points. The central feature of about 50$\mu$m width is an artefact of smoothing in this region of increased noise.} \label{fig:curvdensplot}
    \end{center}
\end{figure*}

\begin{figure}
    \begin{center}
    \includegraphics[width=4.5in]{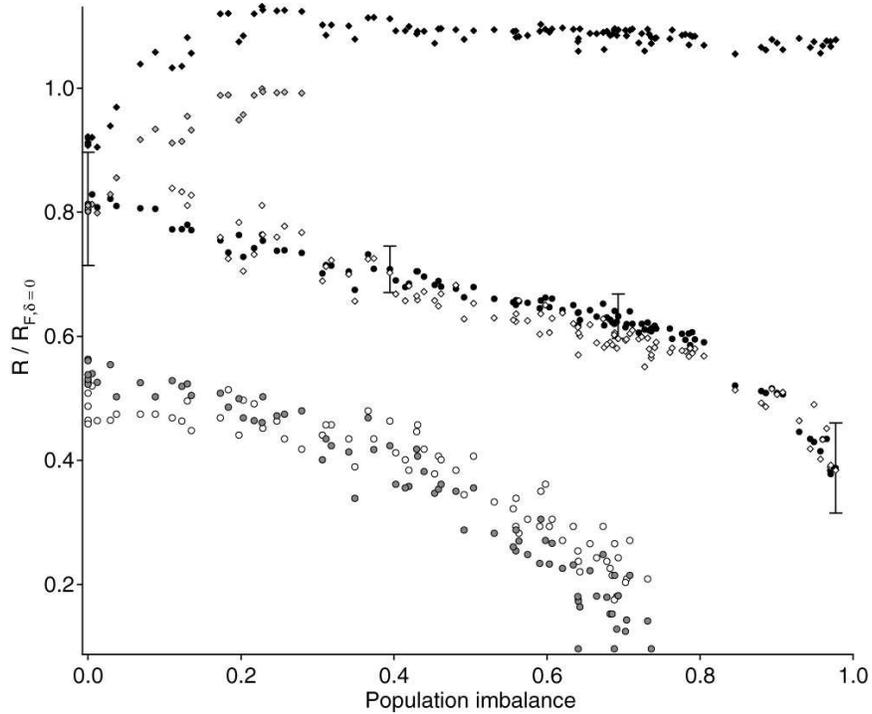}
    \caption[Title]{Outer radii of the two cloud
profiles and condensate radius versus population imbalance. Data
obtained from the majority (minority) cloud are shown as diamonds
(circles). The outer radii of the clouds (black) are determined
from Thomas-Fermi fits to the profiles' wings, where the results of a zero-temperature and a finite temperature fit were averaged. For the minority cloud, the representative error bars indicate the difference
between these two results. The position of the "bulge" in the majority
profile (white diamonds) naturally follows the outer minority
radius. The condensate radius is defined as the position of the
"kink" in the minority profiles. It was obtained by a) fitting an
increasing portion of the minority wings until a significant
increase in $\chi^2$ was observed (grey circles), and b) the
position of the minimum in the profile's derivative (white
circles). All sizes are scaled by the Fermi-radius of a
non-interacting equal mixture. The minority radii were adjusted
for the observed hydrodynamic expansion (expansion factor 11.0).
The non-interacting wings of the majority cloud expand
ballistically (expansion factor 9.7), as long as they are found a
factor $11/9.7 = 1.13$ further out than the minority radius. For small
imbalances ($\delta < 20\%$), also the majority wing's expansion
will be affected by collisions. The grey diamonds give the
majority cloud's outer radius if hydrodynamic expansion is
assumed.} \label{fig:radii}
    \end{center}
\end{figure}

\section*{Supplementary Methods}

\section*{Hydrodynamic vs. ballistic expansion}

A non-interacting cloud of atoms simply expands ballistically from a trap. However, strongly interacting equal Fermi mixtures, above and below the phase transition, are collisionally dense and therefore expand according to hydrodynamic scaling laws~\cite{ohar02science,meno02,cast04scal}. These scaling laws only depend on the equation of state of the gas, $\epsilon \propto n^\gamma$, with $\gamma = 1$ for the BEC-side, $\gamma = 2/3$ for resonance (a direct consequence of unitarity) and $\gamma = 2/3$ for the BCS-side, away from resonance. In an unequal spin mixture of fermions, the expansion does not follow a simple scaling law. The minority cloud is always in contact with majority atoms and thus strongly interacting throughout the expansion, which is therefore hydrodynamic. The excess atoms in the wings of the larger cloud are non-interacting and will expand ballistically, as we have checked experimentally.
The absorption images after expansion are taken along the axial direction of the trap (the direction of the optical trapping beam). In order to compare the expanded cloud sizes to the in-trap Fermi radii of non-interacting clouds (see Fig. 2 and Fig. S2 above) we scale the majority cloud with the ballistic factor for the radial direction
$$\sqrt{\cosh^2\left(2\pi \nu_z t / \sqrt{2}\right) + (\sqrt{2} \nu_r / \nu_z)^2 \sinh^2\left(2\pi \nu_z t / \sqrt{2}\right)},$$
where $t$ is the expansion time and $\nu_z/\sqrt{2}$ gives the radial anti-trapping curvature of the magnetic saddle-point potential.
The scaling factor for the hydrodynamic expansion of an equal mixture is given by the solution to a differential equation~\cite{meno02,cast04scal}. A priori, the minority cloud in unequal mixtures could expand with a different scaling, since the equation of state now depends on {\it two} densities. However, by imaging the cloud in trap and at different times during expansion, we found that the minority cloud's expansion is very well described by the scaling law for an equal mixture. In particular, the aspect ratio of the minority cloud did not change as a function of population imbalance (within our experimental error of 5\%), and was equal to that of a balanced mixture.

For the data on resonance in Figs.~3, S1 and S2, which were obtained after 11 ms expansion out of a trap with radial (axial) frequency of $\nu_r = 113(10)\,\rm  Hz$ ($\nu_z = 22.8(0.2)\,\rm Hz$), the ballistic (hydrodynamic) expansion factor for the radial direction is 9.7 (11.0).

\section*{Supplementary Discussion}

\section*{Signature of the condensate}

Fig.~S1 demonstrates that on resonance, the condensate is visible not only in the minority component, but also in the larger cloud as a small change in the profile's curvature. In the condensate region, the majority profile is slightly depleted when compared to the shape of a normal Fermi cloud. This effect is still significant on the BCS-side (see
Fig.~1): Although here, the condensate is less visible in the smaller component than on resonance, the larger cloud's central depletion still produces a clear dip in the difference profile.

\section*{Radii in the unequal Fermi mixture}

Fig. ~S2 shows the outer radii of the majority and
minority cloud, together with the condensate radius (on
resonance, for the deepest evaporation compatible with constant total atom number versus imbalance). As was the case
for the phase transition at finite temperature, the outer cloud
sizes change smoothly with imbalance. No drastic change is seen at
the critical population imbalance. The radii are obtained by fitting the profiles' wings to the Thomas-Fermi expression for the radial column density $n(r)$:
$$n(r) = n_0 \frac{{\rm Li}_2\left(-\lambda^{1 - r^2/R^2}\right)}{{\rm Li}_2\left(-\lambda\right)},$$
with the central column density $n_0$, the fugacity $\lambda$ and the Thomas-Fermi radius $R$ as the free parameters. ${\rm Li}_2(x)$ is the Dilogarithm. The zero-temperature expression reduces to $n(r) = n_0 (1 - r^2 / R^2)^2$.

\section*{Lower and upper bounds for the critical chemical potential difference at $\delta_c$}

For the clouds at the critical imbalance $\delta_c$, we now want to extract a lower and upper bound for the difference in chemical potentials $\delta\mu_c$ of the majority and minority component. This difference allows us to conclude that BCS-type superfluidity with imbalanced densities is not possible.

The chemical potential difference $\delta\mu \equiv 2 h = (\mu_\uparrow -
\mu_\downarrow)$ measures the energy cost, relative to $\mu =
(\mu_{\uparrow} + \mu_{\downarrow}) /2$, to add a particle to
the cloud of excess fermions. $\Delta$, the pairing gap, is
the energy cost for this additional majority particle to enter the
superfluid. Both the critical temperature $T_C$ and the critical chemical potential difference $\delta\mu_c$ provide a measure of the superfluid gap: The superfluid can be either destroyed by raising
the temperature or by increasing the population imbalance. If $h_c \equiv \delta\mu_c /2
< \Delta$, excess atoms will always stay outside the superfluid,
in the phase separated normal state. For $h_c > \Delta$, excess
atoms can enter the superfluid for $h_c > h > \Delta$. Hence,
superfluidity with unequal densities, if allowed via $h_c > \Delta$, would be favored at
large population imbalance, contrary to the interpretation
in~\cite{part06phaseSUP}, where such a state was proposed for small
population imbalance. A recent Monte-Carlo calculation~\cite{carl05} for the Clogston limit on resonance gives $h_c = 1.00(5)\Delta = 0.50(5) E_F$ and can thus not decide on the question of superfluidity with imbalanced densities.

We can attempt to extract the chemical
potential from the cloud sizes $R_{\uparrow,\downarrow}$ - taking into account hydrodynamic expansion
for the minority cloud and ballistic expansion for the excess
fermions. For the majority cloud, we find $\mu_{c,\uparrow} =
1/2 m \omega_r^2 R_\uparrow^2 = 1.21(6) E_F$.
For the minority cloud, we find $1/2 m \omega_r^2 R_\downarrow^2 = 0.39(10) E_F$.
Throughout the smaller cloud, minority atoms are always strongly attracted by majority atoms. This strong attractive interaction likely reduces their chemical potential from the above upper limit.
The difference of the chemical potentials $\delta\mu_c \equiv 2 h_c$ is thus given by $h_c = (\mu_{c,\uparrow} - \mu_{c,\downarrow})/2\ge 0.41(6) E_F = 0.51 \mu$, our lower bound.
Another condition on $h_c$ concerns whether the normal state can be mixed, $h_c < \mu$, (minority and majority atoms in the same spatial region) or whether the normal state is always completely polarized $h_c > \mu$. Our observation of the mixed region in Fig.~1 immediately results in $h_c < \mu$, the upper bound.

On resonance, $\Delta = 1.16 \mu$ in BCS-theory, while a recent Monte-Carlo study~\cite{carl05} obtains $\Delta = 1.2 \mu$. If $\Delta > \mu$ holds true, our finding of the upper bound on $h_c$ would imply $h_c < \Delta$ and hence would exclude a superfluid with unequal spin densities (at least on the basis of BCS-theory, see~\cite{ho06homo} for a recent suggestion which goes beyond BCS).\\

\section*{Supplementary Notes}

\end{document}